\documentclass[sigconf,nonacm]{acmart}
\usepackage{amsmath,amsfonts}
\usepackage{algorithm}
\usepackage{algorithmic}
\usepackage{graphicx}
\usepackage{textcomp}
\usepackage{xcolor}
\usepackage{amsmath}
\usepackage{subcaption}
\usepackage{xspace}
\usepackage{hyperref}
\usepackage{listings}
\usepackage{booktabs}
\usepackage{tabularx}
\usepackage{multirow}
\usepackage{booktabs}
\usepackage{colortbl}
\usepackage{ctable}
\usepackage{natbib}
\usepackage{comment}
\usepackage{listings}

\newcommand{\squishlist}{
   \begin{list}{$\bullet$}{%
        \setlength{\itemsep}{0pt}%
        \setlength{\parsep}{0pt}%
        \setlength{\topsep}{0pt}%
        \setlength{\partopsep}{0pt}%
        \setlength{\listparindent}{-2pt}%
        \setlength{\itemindent}{-5pt}%
        \setlength{\leftmargin}{1.2em}%
        \setlength{\labelwidth}{0em}%
        \setlength{\labelsep}{0.5em}%
    }
}
\newcommand{\squishend}{
    \end{list}  }

\AtBeginDocument{%
  \providecommand\BibTeX{{%
    \normalfont B\kern-0.5em{\scshape i\kern-0.25em b}\kern-0.8em\TeX}}}

\usepackage[section]{placeins}

\newcommand{\nickname}{VeriAssist} 
\usepackage{tablefootnote}
\usepackage{threeparttable}

\begin{document}

\title{Towards LLM-Powered Verilog RTL Assistant:
Self-Verification and Self-Correction}

\author{
    Hanxian Huang$^{1}$, Zhenghan Lin$^{2}$, Zixuan Wang$^{1}$, Xin Chen$^{3}$, Ke Ding$^{3}$, Jishen Zhao$^{1}$ \\
    $^{1}$University of California San Diego, 
    $^{2}$University of California Berkeley \\ 
    $^{3}$Applied ML Group, Intel Corp. \\
    \texttt{$^{1}$\{hah008, ziw002, jzhao\}@ucsd.edu,} \\
    \texttt{$^{2}$zhenghan\_lin@berkeley.edu,}\\
    \texttt{$^{3}$\{xin.chen, ke.ding\}@intel.com} 
}

\renewcommand{\shortauthors}{Hanxian Huang et al.}

\begin{abstract}
We explore the use of Large Language Models (LLMs) to generate high-quality Register-Transfer Level (RTL) code with minimal human interference.
%
%
The traditional RTL design workflow requires human experts to manually write high-quality RTL code, which is time-consuming and error-prone.
%
With the help of emerging LLMs, developers can describe their requirements to LLMs which then generate corresponding code in Python, C, Java, and more.
%
Adopting LLMs to generate RTL design in hardware description languages is not trivial, given the complex nature of hardware design and the generated design has to meet the timing and physical constraints.
%
%

We propose \textit{\nickname{}}, an LLM-powered programming assistant for Verilog RTL design workflow.
\nickname{} takes RTL design descriptions as input and generates high-quality RTL code with corresponding test benches.
%
%
{\nickname{}} enables the LLM to self-correct and self-verify the generated code by adopting an automatic prompting system and integrating RTL simulator in the code generation loop.
To generate an RTL design, {\nickname{}} first generates the initial RTL code and corresponding test benches, followed by a self-verification step that walks through the code with test cases to reason the code behavior at different time steps, and finally it self-corrects the code by reading the compilation and simulation results and generating final RTL code that fixes errors in compilation and simulation.
This design fully leverages the LLMs' capabilities on multi-turn interaction and chain-of-thought reasoning to improve the quality of the generated code.
We evaluate {\nickname{}} with various benchmark suites and find it significantly improves both syntax and functionality correctness over existing LLM implementations, thus minimizing human intervention and making RTL design more accessible to novice designers.


\end{abstract}

\maketitle

\section{introduction}\label{sec:intro}

Digital hardware design often requires engineers to implement code in hardware description languages (HDLs) such as Verilog and VHDL to define the architecture and functionality of the hardware.
Such hardware design workflows with HDLs are time-consuming and error-prone~\cite{dessouky2019hardfails,ahmad2024hardware,rtl-repair}.
%
%
%
The industry and academia have been improving the electronic design automation (EDA) frameworks to simplify the HDL workflow and improve the design quality~\cite{lavagno2018eda,scheffer2018eda}.
%
%
One such example is the high-level synthesis which enables developers to describe hardware design in high-level programming languages (PLs) such as C and C++ instead of HDLs, thus improving the programmability while at the cost of compromising hardware efficiency~\cite{wakabayashi2000c,nane2015survey}.

Recent advancements in Large Language Models (LLMs), present a promising opportunity to enhance EDA without sacrificing hardware efficiency.
LLMs are pre-trained on large scales of natural or structured language (like PLs) corpora, to establish their foundational capability of generating a sequence of coherent text~\cite{zhao2023survey, zan2022large}.
%
%
Recent works have explored leveraging LLMs to improve the hardware design:
\citet{pearce2020dave} and \citet{thakur2023benchmarking} propose to fine-tune open-source LLMs like CodeGen~\cite{nijkamp2022codegen} to generate Verilog code for target designs.
RTLLM~\cite{lu2024rtllm} and VerilogEval~\cite{liu2023verilogeval} further introduce larger-scale open-source benchmarks to evaluate the RTL code generated by LLMs.
ChipNeoMo~\cite{liu2023chipnemo} trains domain-adaptive models for EDA tasks on a dataset of $\sim 24$ billion tokens with thousands of GPU hours.  

\begin{figure}[t]
    \centering
    \includegraphics[width=\linewidth]{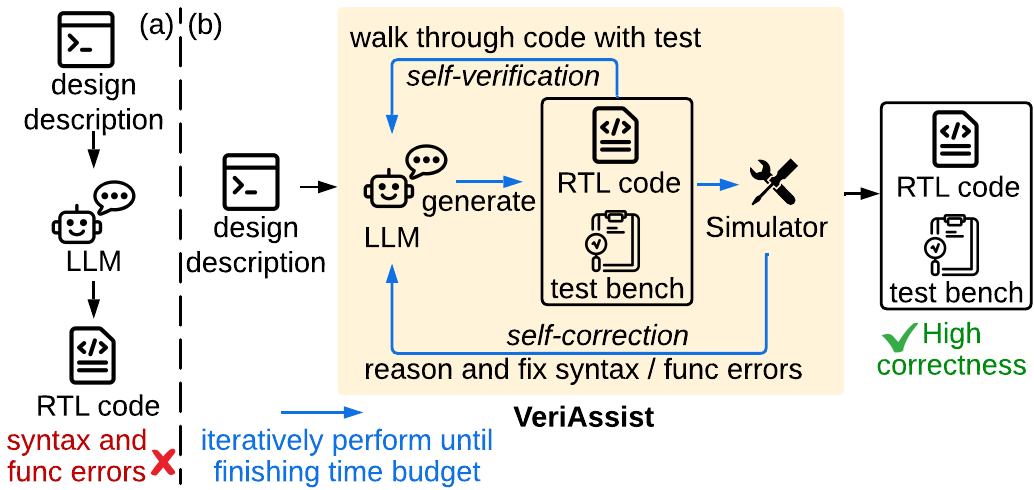}
    \caption{Comparison of workflows between (a) the conventional method of adopting LLMs for RTL code generation and (b) our designed \nickname{}.}
    \label{fig:overview}

\end{figure}

%
We observed these prior works did not fully leverage LLMs' capability of the multi-turn interaction and chain-of-thought~\cite{wei2022chain} (i.e., solve complex tasks step by step).
As a result, they suffer a low RTL design quality, e.g., GPT-4~\cite{achiam2023gpt} achieves a low functionality pass rate of $\sim 60\%$ with simple design tasks~\cite{lu2024rtllm,liu2023verilogeval}. 
We further observed that prior works prompt (i.e., design input text to elicit desired responses) or fine-tune the general-purpose LLMs, which are pre-trained in high-level languages instead of HDLs.
%
%
However, HDLs are designed specifically to describe the logic and architecture of digital hardware at a low level, focusing on the flow of data between registers and the timing of operations, which differs significantly from high-level PLs.
Adopting general LLMs for code generates low-accurate RTL code that cannot fulfill the timing restrictions and thus leads to wrong designs.

In this paper, we incorporate LLM's promising abilities in interactive learning and chain-of-thought reasoning, by mimicking human designer's behaviors to better assist RTL design problems.
Similar to the human design process, instead of generating RTL code in one go, our proposed \nickname{} adopts a multi-turn generation process, as shown in \autoref{fig:overview}.
It starts with an initial prompt based on the digital design task description (Section~\ref{sec:init}), enabling \nickname{} to comprehend the task requirements and develop a step-by-step solution plan.
Then, \nickname{} iteratively performs the following three major steps.
(1) Generating RTL code according to its understanding of the task and the step-by-step plan.
(2) Generating a test bench with test cases if they are not provided by the design specification.
%
%
This step enhances the RTL design through the self-verification process, rather than a final verification.
This involves walking through the generated code with test cases and time settings, analyzing code behavior considering time constraints, and incrementally refining the code based on the analysis (Section~\ref{sec:self-verification}).
(3) If a valid test bench is available, \nickname{} will test the generated RTL code, gather feedback from the simulator, and perform self-correction by identifying and rectifying compilation errors and functional bugs (Section~\ref{sec:self-correction}).
Either the generated RTL code passes the valid test bench or the generation time budget is finished, \nickname{} will stop the generation and provide the best-generated code during this process, suggesting RTL code and test code sketches to assist in the RTL design.

In summary, we make the following contributions:
\squishlist
\item We introduce \nickname{}, a multi-turn Verilog RTL code design assistant that suggests high-quality RTL code with an average pass@5 score of 72.3\%, along with corresponding test benches, demonstrating \nickname{}'s potential to reduce the need for human intervention, making RTL design more accessible to novices.
\item We implement \nickname{} with self-correction capability by refining code based on the simulation feedback. Meanwhile, \nickname{} incorporates self-verification by reasoning the code behavior with test cases considering timing constraints. Our findings indicate that the proposed process of generating test benches and self-code walk-throughs significantly improves the RTL code quality. 
\item Our evaluations across various benchmarks demonstrate the effectiveness of \nickname{}, improving the functionality pass rate by up to 10.4\% on the pass@5 score, and achieving comparable or even better RTL code performance than designer reference code. Our results show that, by better leveraging existing well-pre-trained LLMs and mimicking human RTL design workflow, \nickname{} outperforms the approaches of training or fine-tuning smaller domain-specific models for RTL design. 
\squishend

\section{Background and Related Work} 
Our design is motivated by the characteristics and challenges of Verilog RTL design and the opportunities presented by LLMs. 

\vspace{3pt}
\noindent\textbf{Complexity of Traditional RTL Design.}
Digital hardware design flows necessitate designers to write code in HDLs to specify hardware architectures and behaviors at a granular level. Furthermore, designers must develop and customize test benches to rigorously verify the correctness of these hardware descriptions, ensuring accurate and reliable functioning. The verification process requires substantial experience to resolve subtle timing issues that might only manifest under certain conditions or specific hardware configurations. With the test benches, designers will need to further refine the designs based on feedback from testing, which is another critical phase that can be iterative and cumbersome. This iterative loop among design, verification, and refinement makes the RTL design process not only challenging but also highly demanding in terms of both time and expertise. The complexity of traditional Verilog RTL design calls for Verilog RTL assistants to relieve the burden of manual design and reduce errors during design. 

\vspace{3pt}
\noindent\textbf{Differences from General-Purpose Programming Languages.} Verilog RTL design languages, differ significantly from general-purpose PLs like Python in both their nature and complexity. 
Python and similar PLs are primarily focused on algorithmic logic and data manipulation, allowing developers to abstract away hardware-specific details. The application of LLMs in programming has been widely explored and has achieved notable success in generating correct code in PLs such as Python, C/C++, and Java~\cite{roziere2023code,li2023starcoder,nijkamp2023codegen2,li2022competition}. However, none of them is tailored for Verilog RTL code. 
RTL languages are designed specifically to describe the behavior and architecture of digital hardware at a low level, emphasizing the flow of data between registers and the timing of operations. This focus on timing and hardware characteristics, such as propagation delays and signal dependencies, adds an extra layer of intricacy that is absent in general-purpose PLs, necessitating significant customization to adapt LLMs for effective use in RTL design. Directly applying or slightly fine-tuning these models on Verilog RTL code suffer from low syntactic and semantic accuracy~\cite{lu2024rtllm,liu2023verilogeval,thakur2023benchmarking,liu2023chipnemo}. 

 \begin{figure}[t]
    \centering
    \includegraphics[width=\linewidth]{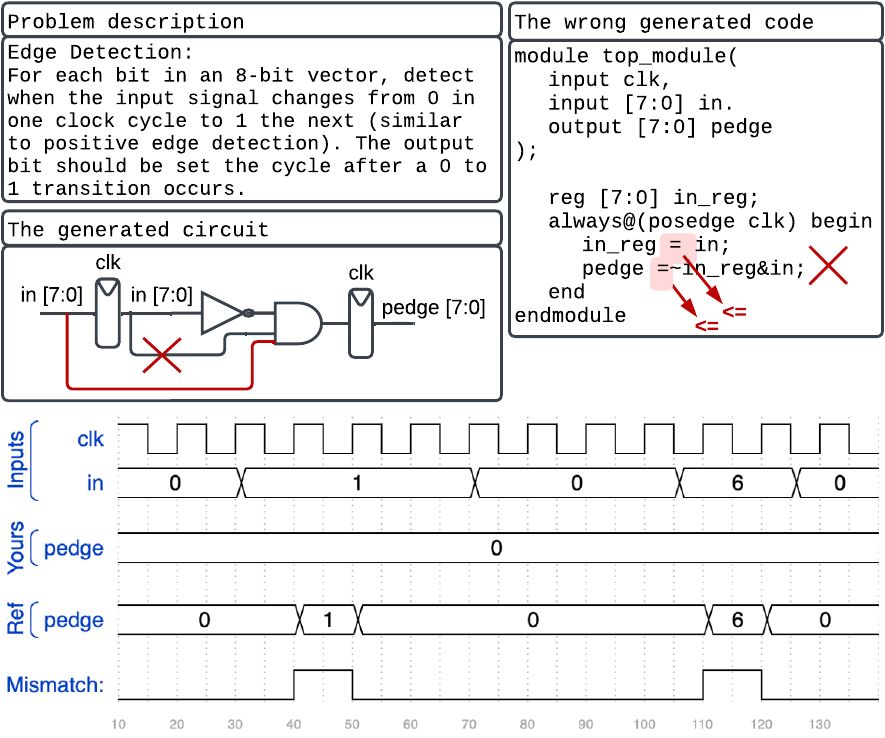}
    \caption{An example of code generated by the conventional method that violates timing constraints and leads to wrong results.}
    \label{fig:motiexample}
\end{figure}

\vspace{3pt}
\noindent\textbf{Related Work on LLMs for EDA.} LLMs are a category of machine learning models that employ transformer architectures~\cite{vaswani2017attention} and are trained on vast language data sets. LLMs operate by examining sequences of input tokens (words or subwords) and predicting the most probable subsequent tokens. The most powerful LLMs, for instance, GPT-4~\cite{achiam2023gpt} and Claude-3~\cite{Claude3} boast hundreds of billions of parameters~\cite{brown2020language} and generalize to a broad range of tasks, e.g., question answering, and code generation. DAVE~\cite{pearce2020dave} first leverages the finetuned GPT-2 model for generating hardware based on design description in natural language, but it does not generalize well to practical Verilog designs. VeriGen~\cite{thakur2023benchmarking} improves upon DAVE by expanding the model size and the size of hardware data sets. RTLLM~\cite{lu2024rtllm} and VerilogEval~\cite{liu2023verilogeval} introduce larger-scale open-source benchmarks for designing RTL generation with natural language, and evaluate prompting and fine-tuned models on them. Chip-Chat~\cite{blocklove2023chip} seeks to evaluate GPT-4 to work with a hardware designer to generate a processor and tape-out. RapidGPT~\cite{RapidGPT} is a new commercial conversational tool for hardware generation. ChatEDA~\cite{wu2024chateda} utilizes LLMs for automating EDA tooling. Moreover works on LLMs for EDA are surveyed in the literature~\cite{zhong2023llm4eda}.

\vspace{3pt}
\noindent\textbf{Limitations of Existing Studies.}
Existing works directly applying or fine-tuning LLMs to RTL design have several limitations. \textit{(L1) Generate RTL code in a single inference and suffer low accuracy.} All of the existing works~\cite{lu2024rtllm,liu2023verilogeval,thakur2023benchmarking} that adopt LLMs on RTL code generation perform the code generation in one go. Although it is feasible to run the models multiple times and pick the best-generated code, this approach overlooks the potential benefits of incremental code refinement within an iterative design and feedback loop. Previous works often generate inaccurate Verilog code, which may result in increased human effort to debug and fix the generated code.  
\textit{(L2) Neglect time constraints.} Unlike conventional PLs, HDLs like Verilog require precise management of timing, synchronization, and parallelism inherent in hardware circuits. As shown in Figure~\ref{fig:motiexample}, the RTL code generated by the conventional method (as shown in Figure~\ref{fig:overview} (a)) neglects the timing features in RTL and leads to a wrong functionality. While Python code executes sequentially, HDLs can describe the events that happen simultaneously, for example, Verilog has combinational logic and synchronized logic allowing non-blocking assignments. Thus, treating RTL using the same approach as Python generates sub-optimal results. 
\textit{(L3) Neglect test bench generation.} The verification of Verilog RTL code is essential for ensuring the functionality and efficiency of designs. 
The test bench itself is also a type of code that can be generated by LLMs. Generating the corresponding test bench as a side task helps the model to understand the design problem and is beneficial for Verilog code refining. Previous works overlook this potential and thus miss the opportunities for code improvement.

\section{\nickname{} Overview}
To address the above challenges and limitations, we design \nickname{}, an LLM-empowered assistant designed to enhance Verilog RTL design. There are several common approaches to adopting LLMs, including prompt engineering~\cite{lu2024rtllm,brown2020language}, fine-tuning~\cite{thakur2023benchmarking,liu2023verilogeval}, and domain-adaptive pre-training~\cite{liu2023chipnemo,roziere2023code,li2023starcoder}. However, fine-tuning and domain-adaptive pre-training are costly, requiring vast domain-specific datasets and substantial training resources. Additionally, their performance hinges on the availability of sufficient high-quality data~\cite{wei2022emergent}. Consequently, we opted for prompt engineering, i.e., structuring the prompts to frame the task so that LLMs can understand and generate the desired output particularly focusing on the RTL design. Note that designing effective prompts and automating the prompting generation is non-trivial. Existing prompt engineering~\cite{lu2024rtllm,liu2023verilogeval} on RTL code generation focuses on crafting high-quality problem descriptions. Yet, the generated results are unsatisfactory with a low pass rate ($\sim 60\%$) compared to other programming languages such as Python ($\sim 80\%$)~\cite{muennighoff2023octopack,huang2023agentcoder,achiam2023gpt}. 

\begin{figure}[t]
    \centering
    \includegraphics[width=0.8\linewidth]{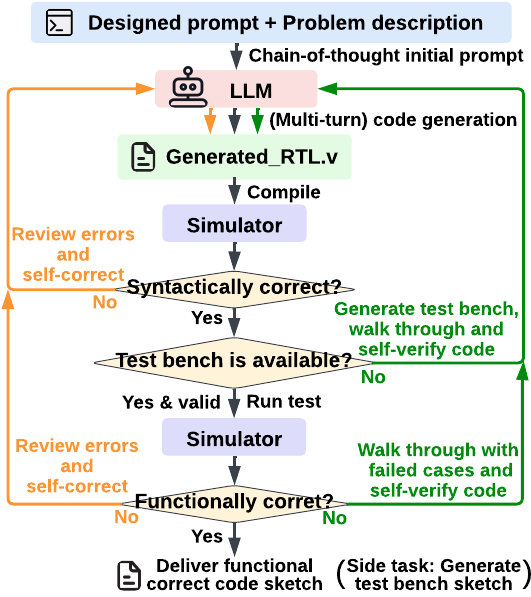}
    \caption{\nickname{} workflow.}
    \label{fig:workflow}
 
\end{figure}

Different from existing studies, \nickname{} is inspired by the human RTL design workflow, where the designers: (1) plan the solution step by step; (2) design based on the planning; (3) based on the RTL code and design requirements, craft a test bench with test cases, walk through the code with test cases across time steps; (4) run the simulation and debug the code based on simulation results; (5) iteratively design and refine. \nickname{} is an automatic prompting system with a similar workflow integrating with the simulator feedback loop. It leverages the code generation, multi-turn interaction, and chain-of-thought reasoning capabilities of LLMs to iteratively generate and optimize RTL code. With well-crafted prompt engineering, \nickname{} delivers effective prompts to LLMs that can lead to more accurate, relevant, and context-aware responses, enabling LLMs to self-verify the code with various test cases, taking timing constraints into account, and self-correct the code by fixing bugs reported by simulator. As shown in Figure\ref{fig:workflow}, the workflow of \nickname{} can be summarized into the following steps:

\textbf{(1) Initial Prompt Preparation}: Given a problem description as input, \nickname{} transforms it into a prompt and feeds it into the LLM. This prompt guides the model to grasp the design requirements and devise a step-by-step plan to address the problem.

\textbf{(2) RTL Code Generation}: The LLM interprets the prompt and generates a corresponding RTL code. In each generation, the model employs in-context learning and chain-of-thought reasoning, allowing it to understand the prompt, reason through the design, and produce code.

\textbf{(3) Syntactical Verification}: The generated RTL code is compiled, and its syntax is checked. If syntax errors are detected, feedback from the simulator’s error logs is used to prompt the LLM model for self-correction. \nickname{} then revises the code iteratively until it passes the syntactical check.

\textbf{(4) Functional Verification}: Once the code passes the syntactical check, it is passed to the functionality check. If a valid test bench is available, \nickname{} runs functional tests to validate the design. Otherwise, \nickname{} is prompted to generate a test bench, walk through the code, and perform self-verification. If functional errors are detected, \nickname{} prompts the model to walk through the code with failed cases and revise the code accordingly.

\textbf{(5) Iterative Design and Refinement}: \nickname{} iteratively performs design and refinement 
through Steps (2)-(4) until finishing time budget or passing the functionality check, and finally outputs the Verilog RTL code and the corresponding test bench sketch.

\section{\nickname{} Design}
The design of \nickname{} includes constructing an initial prompt from the problem description, followed by prompts for self-verification and self-correction. The principles of prompt designing are guided by the goal of mimicking human designer behaviors—implementing a step-by-step methodology that encompasses planning, designing, testing, and debugging. With the well-crafted prompts, \nickname{} is designed as an automatic prompting system integrating with a simulator to facilitate the prompt-generate-feedback-revise loop.

\subsection{Problem Description and Initial Prompt}\label{sec:init}
\begin{figure}[t]
    \centering
    \includegraphics[width=\linewidth]{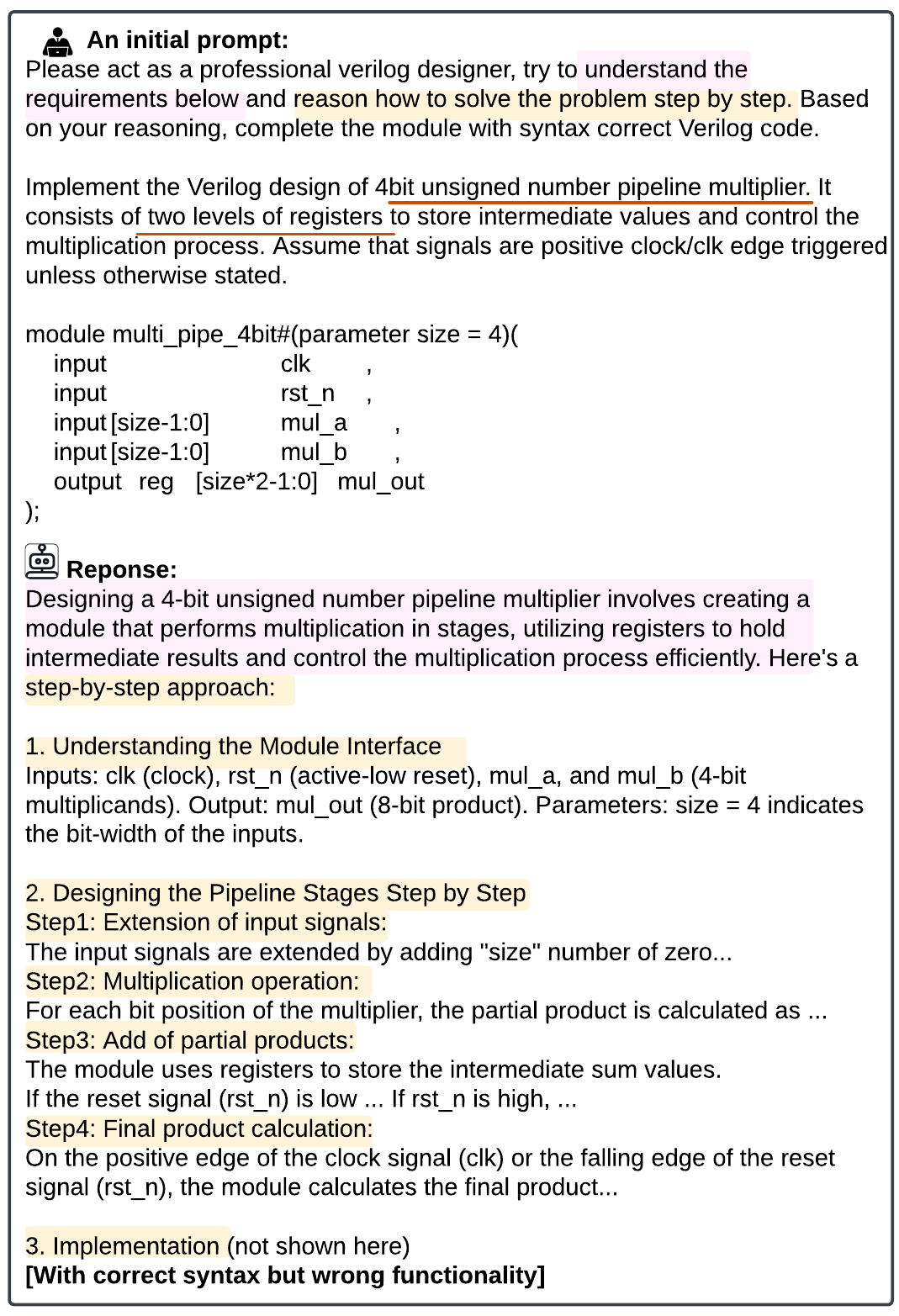}
    \caption{An example of an initial prompt.}
    \label{fig:inti_prompt}
 
\end{figure}

The input of \nickname{} is a problem description, which includes a natural language description of the task, signals, triggering condition, as well as the module header, and input / output definition in HDL. Some problem descriptions also include implementation requirements such as a specific algorithm, the usage of the pipeline and registers, etc. Besides the problem description, we include a \textit{system prompt} at the beginning of the model input. It provides more contextual information and defines the scope of the LLM's capabilities to enable it to focus on the Verilog RTL design domain. The system prompt is fixed for all the design tasks. The system prompt is concatenated with the problem description and sent to the LLM for inference. Figure~\ref{fig:inti_prompt} shows an example of an initial input prompt, where the first paragraph is the system prompt. The system prompt (1) specifies the role of the model as a ``professional Verilog designer'', (2) guides the model first to understand the task description, (3) then instructs the model to generate step-by-step planning before code generation. Compared to conventional methods which directly ask the model to generate code based on the problem description, our method leverages a divide-and-conquer strategy, allowing the LLM to effectively manage complex designs by breaking them into more manageable sub-tasks: understanding, planning, and generation.

An example of the model response is also shown in Figure~\ref{fig:inti_prompt}: (1) By taking the designed initial prompt, the LLM model first analyzes the task requirement of a 4-bit pipeline multiplier and the implementation requirement of using two-level registers to hold intermediate results (marked in pink). 
(2) It then conducts the step-by-step approach according to the requirements (marked in orange), which includes extending input signals, multiplying each bit, adding partial products, and generating final results in the pipeline. (3) Based on the planning, the model generates some initial output code.  However, it is challenging for existing LLMs to develop an entirely correct design in one go~\cite{lu2024rtllm,liu2023verilogeval,chen2021evaluating}. To resolve this issue and improve the generated code, we propose further integrating self-verification and self-correction through an automatic multi-turn prompt-generate-feedback-revise loop system in \nickname{}.

\begin{figure}[t]
    \centering
    \includegraphics[width=\linewidth]{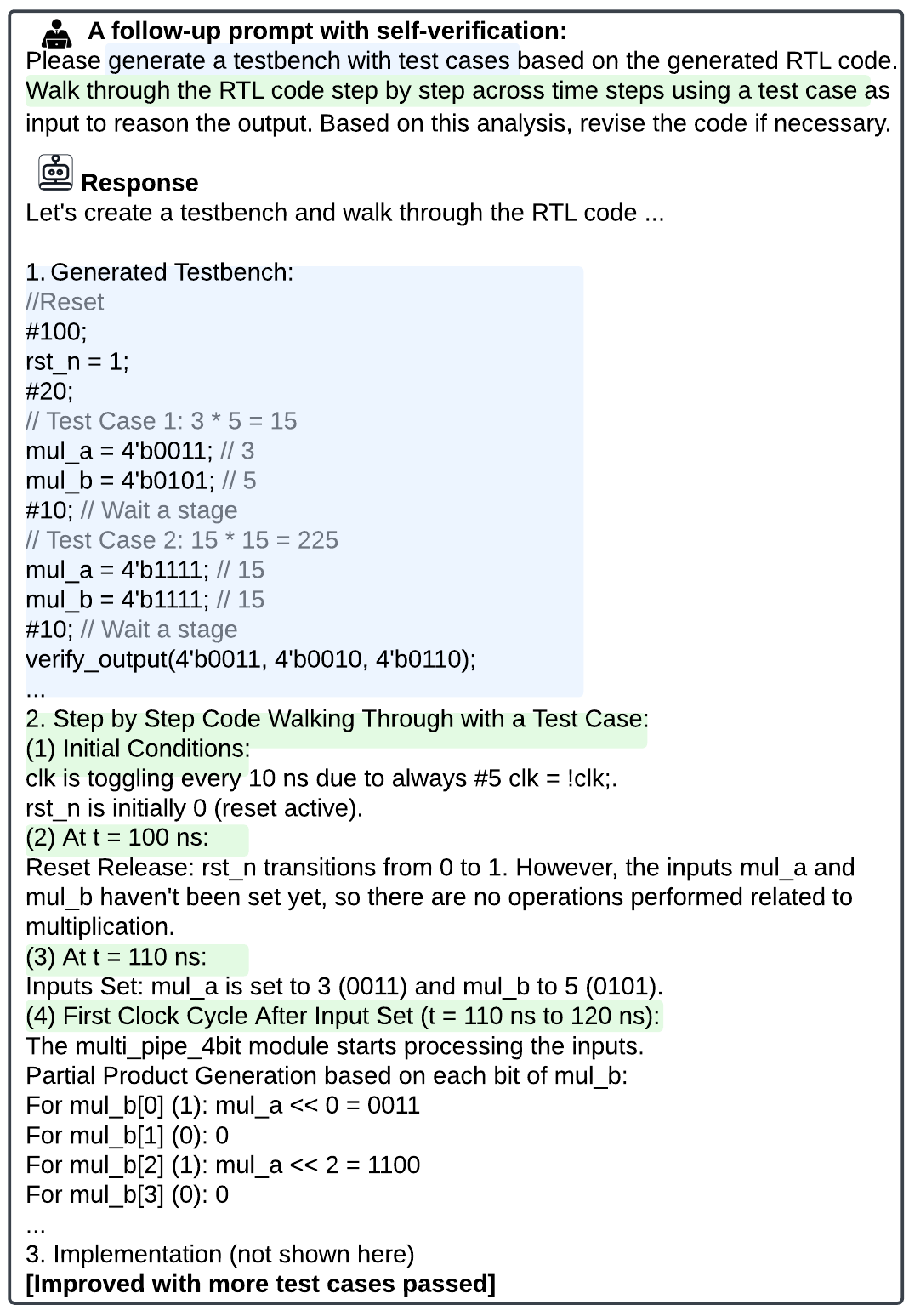}
    \caption{An example of a self-verification prompt.}
    \label{fig:self-veri}
  
\end{figure}

\subsection{Self-Verification}\label{sec:self-verification}
Verification is a critical phase in RTL design to promise design quality. In practice, the test bench and test cases are usually unavailable during design, necessitating designers to carefully craft them. With the test cases, the designers could deductively reason the existing design, and understand better the RTL code behavior to debug and refine the code. 
Similarly, in \nickname{}, when a valid test bench is unavailable, it automatically adopts the \textit{self-verification prompt} which instructs the model to generate the test bench with test cases and timing settings according to the design requirements, and to walk through the code considering the time constraints. When the test bench is available, the \textit{self-verification prompt} will directly guide the model to trace the code with test cases across time steps and revise the code accordingly. 
Note that in a traditional sense, LLMs do not have explicit execution capabilities like a simulator. 
However, we include this code walk-through process to provide more relevant context, facilitating the model to understand the code logic and pattern to refine and debug the code in the next iteration of code generation.

\autoref{fig:self-veri} shows an example of a follow-up prompt with self-verification as well as the model response. In the example, following the prompt, the model first generates the test bench with specific test cases and time settings (marked in blue). Then, it walks through the test code and reasons how signals and registers change across various time steps (marked in green). Based on the code walk-through context, the model refines the RTL code, which can pass more test cases in our study. As we will show later in Section~\ref{sec:effect_self_veri}, generating a test bench, even as a side task, is beneficial for the model to understand the design and requirements, contributing to RTL code optimization. While integrating self-verification helps LLMs comprehend the code logic, there is no guarantee that the generated test bench will be fully correct, and the LLMs themselves cannot physically execute the test code or guarantee correct outputs. 
Therefore, we include the simulator in the design loop to provide real feedback from the execution with LLMs to further revise the code.

 \begin{lstlisting}[caption=An example of syntax error report, label={lst:logSyntax}]
 adder.v:17: Unknown module type: sum
 \end{lstlisting}

 \begin{lstlisting}[caption=An example of functionality error report, label={lst:LogFunc}]
 Test failed: input 4, 5; expect 9, but got 1
 \end{lstlisting}

\subsection{Self-Correction}\label{sec:self-correction}
Providing compilation and execution feedback to  LLMs has been explored in related works on Python and C code generation~\cite{qiao2023taskweaver, chen2023teaching}. We adopt a similar idea to provide LLMs with feedback from the simulation logs, including syntax errors and failures when running the test benches. The simulator receives the generated code, compiles it, and executes the test bench. If the compilation fails, a compilation log will be collected, which pinpoints the location of syntactical errors (e.g., \autoref{lst:logSyntax}). A \textit{self-correction prompt} will be automatically generated to direct the model to fix the syntax errors by paying more attention to the location where the errors happen. If the compilation is successful, \nickname{} will run the test benches and collect a report on any failed test cases (e.g., \autoref{lst:LogFunc}). A prompt will be generated to instruct the model to walk through the failed test cases, analyze the potential reasons, and then regenerate the RTL code based on the analysis, similar to the self-verification process. 

\begin{figure}[t]
    \centering
    \includegraphics[width=\linewidth]{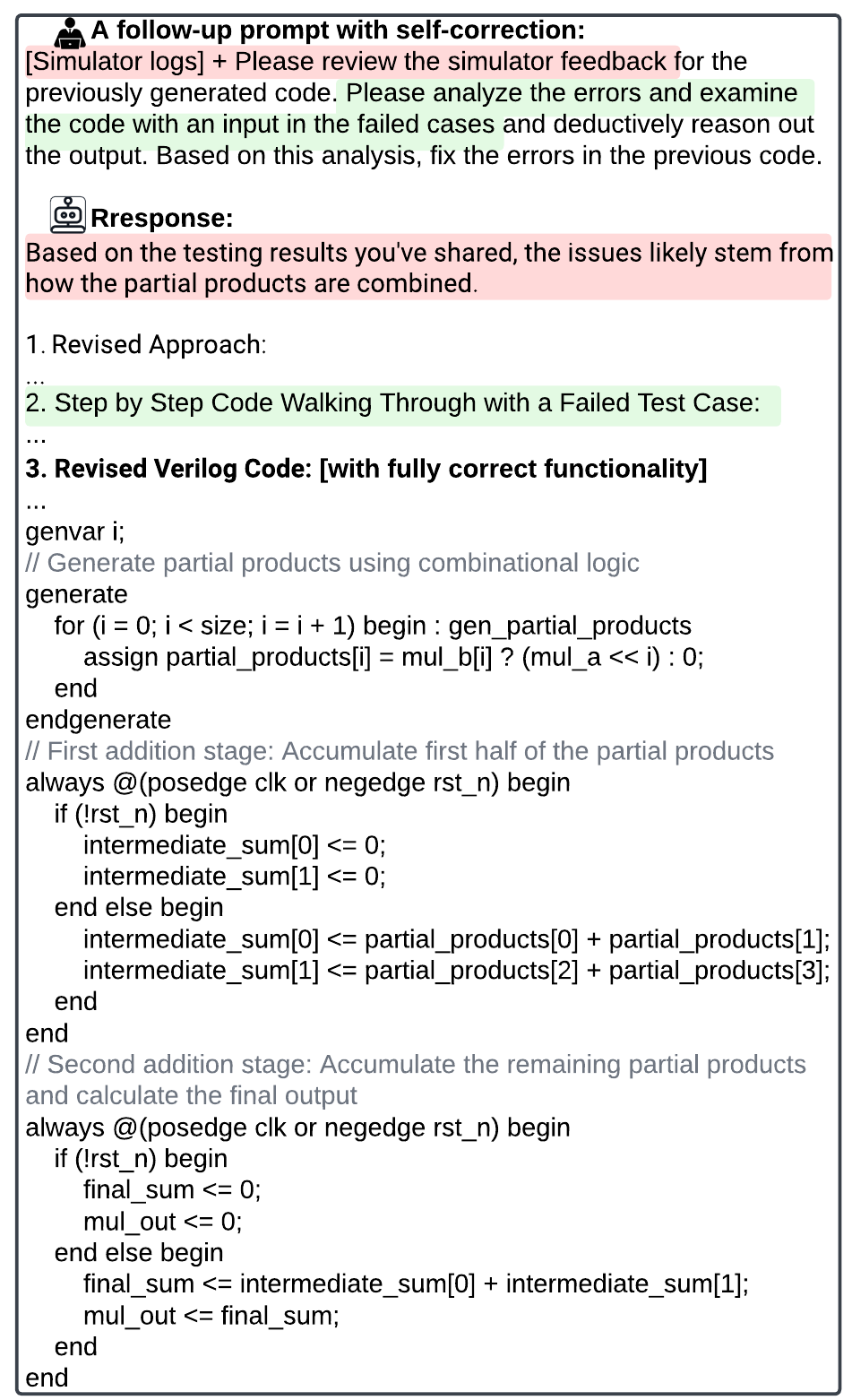}
    \caption{An example of a self-correction prompt.}
    \label{fig:self-correction}
 
\end{figure}

Figure~\ref{fig:self-correction} shows an example of a self-correction prompt where the model first infers the possible reason for the failure based on the simulation logs (marked in red), specifically pointing out the step that probably caused that failure. It then provides a step-by-step revised plan and walks through the failed cases (marked in green, but the detailed plan is not shown here). Based on the real simulation feedback and the self-correction context, the model revises its previously generated code. As shown in the example, the model finally generates an RTL code that correctly implements the multiplier in a pipelined manner with two levels of registers as required.

Putting everything together, \nickname{} performs automatic prompt engineering, interacts with the LLM, calls the simulator for compilation, and collects feedback from compilation logs to achieve iteratively RTL code designing and refining, as shown in Figure~\ref{fig:workflow}.


\section{Experiment}\label{sec:exp}

\subsection{Implementation}
Our implementation of \nickname{} integrates Verilog simulator with several state-of-the-art (SOTA) conversational LLMs. \nickname{} is implemented in Python and manages function calls to simulator and extracts the generated code from the LLM responses. In this paper, we evaluate \nickname{} using GPT-4, GPT-3.5, and Claude-3 Sonnet and call them by their respective model APIs~\cite{OpenAIAPI,ClaudeAPI}. The \nickname{} framework is designed to be orthogonal to the underlying models,  provided that their pre-trained versions are available or accessible via API. For simulation, \nickname{} employs Icarus Verilog (iVerilog)~\cite{iverilog} since it is open source and requires no setup beyond providing a Verilog module and its test bench, introducing very low overhead - it takes on average $\sim$1 second to compile and run a test bench. We set the time budget of generation iteration as 20. On average, the total overhead for \nickname{} to iteratively generate prompts, call model to perform inferences, and run simulations on iVerilog is $\sim$2 minutes per RTL design task.  

\subsection{Experiment Setup}
\noindent\textbf{Benchmarks:}
In evaluation, we utilize two benchmark datasets, covering various complexities and scales, encompassing all common bit widths from 4 bits to 64 bits, and varied implementation requirements. 
(1) RTLLM benchmark~\cite{lu2024rtllm}. This dataset comprises 29 designs, including 11 arithmetic and 18 logic designs.
(2) VerilogEval benchmark~\cite{liu2023verilogeval}. This benchmark includes two parts: VerilogEval-Human, consisting of 156 designs with human-generated problem descriptions, and VerilogEval-Machine, consisting of 143 designs with machine-generated problem descriptions. Although both benchmarks provide valid test benches, we still set up \nickname{} to perform both self-verification and self-correction by default.

\noindent\textbf{Baselines:}
We compare \nickname{} with various models, including both closed-source and open-source models:\\
(1) Closed-source models: 
\squishlist
\item Commercial Models: We include GPT-4, GPT-3.5, and Claude-3 Sonnet, which are pre-trained for general auto-aggressive text generation but not specifically for RTL code generation.
\item Domain-Adaptive Pre-Training Model: The ChipNeMo model \cite{liu2023chipnemo} trains the advanced Llama2 model \cite{touvron2023llama} on 24 billion tokens of chip design documentation and code, and offers model variants with different sizes, e.g., 13 billion and 70 billion parameters. 
\item Fine-tuned Model: The verilog-sft-16B \cite{liu2023verilogeval} model, which is a fine-tuned version of the CodeGen model \cite{nijkamp2022codegen} with Verilog data. But this particular model is not publicly available. 
\squishend
(2) Open-source models:
\squishlist
\item Pre-trained Model: CodeGen2-16B is a general academic model~\cite{nijkamp2023codegen2} pre-trained on Python but has not been fine-tuned for Verilog.
\item Fine-tuned Model: CodeGen-Verilog-16B is introduced in Thakur et al. \cite{thakur2023benchmarking}, which fine-tunes the CodeGen model \cite{nijkamp2022codegen} with Verilog data sourced from GitHub and the Verilog Books Corpus.
\squishend

\noindent\textbf{Metrics:}
Following recent related studies \cite{lu2024rtllm,chen2021evaluating,liu2023verilogeval}, we employ the pass@$k$ metric to directly assess code functional correctness:
\begin{equation}
    pass@k = \mathbb{E}_{\text{Problems}} \left[ 1 - \frac{{n-c \choose k}}{{n \choose k}} \right]
\end{equation}

\noindent{where $n$ is the total number of trials for each instruction and $c$ is the number of correct code generations for a task. We set $n = 20$ in our experiment. If any code in the $k$ trials could pass the test, then this task is considered to be addressed and the pass@$k$ metric reflects the estimated proportion of design tasks that could be solved.}

Following the benchmark settings, we evaluate across various settings of $k \in [1,5,10]$ for the VerilogEval benchmark. We measure both the syntax pass rate, assessing the success of compilation, and the functionality pass rate, evaluating the success of passing test bench, under the pass@5 setting for the RTLLM benchmark. 

Additionally, we perform a performance analysis based on an FPGA, considering timing, power, and utilization.


\subsection{Results}

\subsubsection{\textbf{Generation Correctness}}
We evaluate the RTL code generation correctness of the \nickname{} against various widely-used code generation LLMs and show the results in Table~\ref{tab:passrate}. In summary, \nickname{} consistently outperforms all the baselines across both benchmarks and achieves significant improvement when adopting SOTA models such as Claude-3, GPT-3.5, and GPT-4. 

The CodeGen2 model, pre-trained on Python corpus, can achieve a $\sim40\%$ pass@5 score on Python code generation task~\cite{nijkamp2023codegen2}. However, its performance on both VerilogEval and RTLLM is less than 10\% for pass@5 scores, highlighting challenges in zero-shot transfer from general PLs like Python to Verilog RTL code generation tasks, given different natures and requirements between Python programming and RTL design. 
In contrast, the fine-tuned CodeGen-Verilog-16B shows substantial improvements over the pre-trained version, demonstrating the benefits of training on a Verilog corpus. 
The verilog-sft-16B model, fine-tuned with more synthetic data, achieves comparable or even better results compared to CodeGen-Verilog-16B. 
ChipNeMo models utilize domain-adaptive pre-training, with the larger 70B model outperforming the smaller 13B version, indicating that scaling up the model size enhances performance. The ChipNeMo-70B model surpasses all the CodeGen model variants, including those with or without fine-tuning.

Claude-3, GPT-3.5 and GPT-4 are well pre-trained commercial models, with potentially over hundreds of billions of parameters~\cite{brown2020language}. 
These models outperform all the smaller models with 13B / 16B / 70B parameters. Conventional methods such as single-iteration inference in VerilogEval-Machine / VerilogEval-Human and simple self-planning scheme in RTLLM achieve only $60.0\%, 43.5\%$ (pass@1) and $65.5\%$ (pass@5) scores with the SOTA model GPT-4, respectively. 
Remarkably, \nickname{} consistently surpasses these traditional generation schemes when employing Claude-3, GPT-3.5, or GPT-4 for RTL code generation. 
Notably, \nickname{} improves 7.5\% and 7.0\% pass@1 scores on VerilogEval-Machine and VerilogEval-Human, and improves 10.4\% on the functionality pass@5 score on RTLLM with a syntax pass rate of 100\% using GPT-4, demonstrating the \nickname{}'s capability in generating high-quality, functionally correct and syntactically accurate Verilog code. Besides, the consistent improvement observed across various LLMs also underscores \nickname{}'s compatibility with different model choices.

\begin{table*}[]
\caption{Pass rate (\%) comparison of RTL code generators on VerilogEval~\cite{liu2023verilogeval} and RTLLM~\cite{lu2024rtllm} benchmarks.}
\label{tab:passrate}
\resizebox{0.95\linewidth}{!}{
\begin{threeparttable}
\begin{tabular}{|c|c|ccc|ccc|cc|}
\hline
     \multirow{2}{*}{Model Type}&   \multirow{2}{*}{Evaluated Model}    & \multicolumn{3}{c|}{VerilogEval-Machine}                            & \multicolumn{3}{c|}{VerilogEval-Human}                              & \multicolumn{2}{c|}{RTLLM\tnote{$\ddag$} pass@5}          \\ \cline{3-10}
       
&  & \multicolumn{1}{c|}{pass@1} & \multicolumn{1}{c|}{pass@5} & pass@10 & \multicolumn{1}{c|}{pass@1} & \multicolumn{1}{c|}{pass@5} & pass@10 & \multicolumn{1}{c|}{Syntax(\%)} & Func(\%) \\ \hline
 
 \multirow{2}{*}{ \begin{tabular}[c]{@{}c@{}}Open-Source\\ Model\end{tabular}}
 
 &CodeGen2-16B~\cite{nijkamp2023codegen2}     & \multicolumn{1}{c|}{5.00}   & \multicolumn{1}{c|}{9.00}   & 13.9    & \multicolumn{1}{c|}{0.90}   & \multicolumn{1}{c|}{4.10}   & 7.25    & \multicolumn{1}{c|}{72.4}       & 6.90     \\  \cline{2-10}

&CodeGen-Verilog-16B~\cite{thakur2023benchmarking} & \multicolumn{1}{c|}{44.0}   & \multicolumn{1}{c|}{52.6}   & 59.2    & \multicolumn{1}{c|}{30.3}   & \multicolumn{1}{c|}{43.9}   & 49.6    & \multicolumn{1}{c|}{86.2}       & 24.1     \\ \hline


 \multirow{6}{*}{ \begin{tabular}[c]{@{}c@{}}Closed-Source\\ Model\end{tabular}} 

&ChipNeMo-13B~\cite{liu2023chipnemo}\tnote{$\dag$}      & \multicolumn{1}{c|}{43.4}   & \multicolumn{1}{c|}{N/A}   & N/A   & \multicolumn{1}{c|}{22.4}   & \multicolumn{1}{c|}{N/A}   & N/A    & \multicolumn{1}{c|}{N/A}       & N/A      \\  \cline{2-10}

&ChipNeMo-70B~\cite{liu2023chipnemo}\tnote{$\dag$}      & \multicolumn{1}{c|}{53.8}   & \multicolumn{1}{c|}{N/A}   & N/A   & \multicolumn{1}{c|}{27.6}   & \multicolumn{1}{c|}{N/A}   & N/A    & \multicolumn{1}{c|}{N/A}       & N/A      \\  \cline{2-10}

&verilog-sft-16B~\cite{liu2023verilogeval}\tnote{$\dag$}  & \multicolumn{1}{c|}{46.2}   & \multicolumn{1}{c|}{67.3}   & 73.7    & \multicolumn{1}{c|}{28.8}   & \multicolumn{1}{c|}{45.9}   & 52.8    & \multicolumn{1}{c|}{N/A}       & N/A     \\  \cline{2-10}
 
&Claude-3~\cite{Claude3}       & \multicolumn{1}{c|}{55.3}   & \multicolumn{1}{c|}{63.8}   & 69.4 & \multicolumn{1}{c|}{34.4}   & \multicolumn{1}{c|}{48.3}   & 53.4    & \multicolumn{1}{c|}{93.1}       & 55.2    \\  \cline{2-10}

&GPT-3.5      & \multicolumn{1}{c|}{46.7}   & \multicolumn{1}{c|}{69.1}   & 74.1    & \multicolumn{1}{c|}{26.7}   & \multicolumn{1}{c|}{45.8}   & 51.7    & \multicolumn{1}{c|}{89.7}       & 37.9     \\\cline{2-10}

&GPT-4        & \multicolumn{1}{c|}{60.0}   & \multicolumn{1}{c|}{70.6}   & 73.5    & \multicolumn{1}{c|}{43.5}   & \multicolumn{1}{c|}{55.8}   & 58.9    & \multicolumn{1}{c|}{\textbf{100}}       & 65.5     \\ \hline\hline

\multirow{6}{*}{ \nickname{}}

& Ours + Claude-3        & \multicolumn{1}{c|}{63.8}   & \multicolumn{1}{c|}{70.4}   & 78.4    & \multicolumn{1}{c|}{41.6}   & \multicolumn{1}{c|}{55.5}   & 62.5    & \multicolumn{1}{c|}{96.6}       & 65.5     \\  
& Improvement $(\Delta)^*$        & \multicolumn{1}{c|}{+8.5}   & \multicolumn{1}{c|}{+6.6}   & +9.0    & \multicolumn{1}{c|}{+7.2}   & \multicolumn{1}{c|}{+7.2}   & +9.1    & \multicolumn{1}{c|}{+3.5}       & +10.3     \\\cline{2-10} 

& Ours + GPT-3.5     & \multicolumn{1}{c|}{55.3}   & \multicolumn{1}{c|}{76.5}   & 80.1    & \multicolumn{1}{c|}{34.4}   & \multicolumn{1}{c|}{51.3}   & 58.9    & \multicolumn{1}{c|}{93.1}       & 48.3     \\ 
& Improvement $(\Delta)^*$         & \multicolumn{1}{c|}{+8.6}   & \multicolumn{1}{c|}{+7.4}   & +6.0    & \multicolumn{1}{c|}{+7.7}   & \multicolumn{1}{c|}{+5.5}   & +7.2    & \multicolumn{1}{c|}{+3.4}       & +10.4     \\\cline{2-10} 

& Ours + GPT-4        & \multicolumn{1}{c|}{\textbf{67.5}}   & \multicolumn{1}{c|}{\textbf{78.3}}   & \textbf{83.2}    & \multicolumn{1}{c|}{\textbf{50.5}}   & \multicolumn{1}{c|}{\textbf{62.8}}   & \textbf{69.2}   & \multicolumn{1}{c|}{\textbf{100}}       & \textbf{75.9 }    \\ 
& Improvement $(\Delta)^*$        & \multicolumn{1}{c|}{+7.5}   & \multicolumn{1}{c|}{+7.7}   & +9.7    & \multicolumn{1}{c|}{+7.0}   & \multicolumn{1}{c|}{+7.0}   & +10.3    & \multicolumn{1}{c|}{0.0}       & +10.4     \\\cline{2-10} 

\hline\hline

\multirow{2}{*}{ \begin{tabular}[c]{@{}c@{}}Ablation\\ Study\end{tabular}}& Self-Verification + GPT-4        & \multicolumn{1}{c|}{63.8}   & \multicolumn{1}{c|}{73.2}   &   78.4  & \multicolumn{1}{c|}{48.3}   & \multicolumn{1}{c|}{58.9}   & 64.7   & \multicolumn{1}{c|}{96.6}       &  69.0    \\ \cline{2-10}
 
& Self-Correction + GPT-4        & \multicolumn{1}{c|}{62.5}   & \multicolumn{1}{c|}{72.2}   &    77.2 & \multicolumn{1}{c|}{47.1}   & \multicolumn{1}{c|}{58.9}   & 66.0 & \multicolumn{1}{c|}{100}       &   69.0  \\ \hline

\end{tabular}
\begin{tablenotes}
    \footnotesize
    \item[$\dag$] The results are referenced from the original papers as the closed-source models are not available.
    \item[$\ddag$] We utilized the RTLLM v1.1 available at \url{https://github.com/hkust-zhiyao/RTLLM.}
    \item[$*$] The improvement is directly compared to the performance of the same model without employing the \nickname{} framework.
  \end{tablenotes}
\end{threeparttable}}
\end{table*}

\subsubsection{\textbf{Performance Analysis}}
We evaluate the RTL code performance by compiling them and analyzing the FPGA resource utilization.
We compile the code using Intel Quartus Prime Pro v23.4, targeting the Intel Agilex-7 FPGA AGIB027R29A1E2VR3.
For each RTL design from the RTLLM benchmark suite~\cite{lu2024rtllm}, we generate a baseline RTL code GPT-4 (following Figure~\ref{fig:overview} (a)), and a code using GPT-4 with our {\nickname} (following Figure~\ref{fig:overview} (b)).
We choose the following metrics reported by Intel Quartus's post-compilation analysis:

\squishlist
    \item \textbf{ALM}: Number of \emph{Adaptive Logic Modules} on device required by the design. The ALM utility can be used to estimate the area.
    \item \textbf{ALUT}: Combinational \emph{Adaptive Look-Up Tables} used for RTL logic.
    \item \textbf{INT}: Number of \emph{Block Interconnects} used in routing.
    \item \textbf{Slack}: Setup time slack. A negative value indicates a setup time violation, and a larger positive value indicates a better setup timing. Cells implemented only by Combinational logic do not have setup/hold time constraints and are marked as N/A. 
    \item \textbf{mPWR}: On-chip power (mW) of the Verilog module estimated by the synthesize tools.
\squishend

\noindent
\autoref{tab:fpga-util} shows our evaluation of FPGA resource utilization where the colored table cell marks the best utilization among three systems, and the ``-'' sign indicates the corresponding RTL code fails to compile or fails to pass the functionality check. All the tests run under the same constraints and optimization settings.
From this result we find {\nickname} improves the generate code's utilization rate by up to 45\% compared to the GPT-4 baseline, achieving comparable or even better performance compared to the designer reference. In summary, {\nickname} significantly improves the RTL code quality compared to baseline, by improving the compilation success rate and FPGA utilization rate.

Note that in the current \nickname{} system, we do not yet incorporate synthesis feedback or prompt the model to optimize the RTL design for runtime performance, power, or area explicitly. It remains challenging since LLM inherently cannot understand the compiler and hardware specifics. Additionally, integrating synthesis feedback into the loop would significantly increase the overhead of the RTL design assistant workflow. We plan to explore the generation of performance-optimized RTL code in our future work.

\begin{table*}
\centering
\caption{FPGA resource utilization.}
\label{tab:fpga-util}
\resizebox{\linewidth}{!}{
\begin{tabular}{|l|rrrrr|rrrrr|rrrrr|}
\FL
	\multirow{2}{*}{\textbf{Design}} & \multicolumn{5}{|c|}{\textbf{Designer Reference~\cite{lu2024rtllm}}} & \multicolumn{5}{|c|}{\textbf{GPT-4 Direct Baseline}} & \multicolumn{5}{|c|}{\textbf{Ours: GPT-4 with {\nickname}}}	\NN \cmidrule(r){2-16}
	~ & \multicolumn{1}{|c|}{\textbf{ALM}} & \multicolumn{1}{|c|}{\textbf{ALUT}} & \multicolumn{1}{|c|}{\textbf{INT}} & \multicolumn{1}{|c|}{\textbf{Slack}} & \multicolumn{1}{|c|}{\textbf{mPWR}} & \multicolumn{1}{|c|}{\textbf{ALM}} & \multicolumn{1}{|c|}{\textbf{ALUT}} & \multicolumn{1}{|c|}{\textbf{INT}} & \multicolumn{1}{|c|}{\textbf{Slack}} & \multicolumn{1}{|c|}{\textbf{mPWR}} & \multicolumn{1}{|c|}{\textbf{ALM}} & \multicolumn{1}{|c|}{\textbf{ALUT}} & \multicolumn{1}{|c|}{\textbf{INT}} & \multicolumn{1}{|c|}{\textbf{Slack}} & \multicolumn{1}{|c|}{\textbf{mPWR}}	\ML
	JC\_counter & 17 & \cellcolor[RGB]{208,240,192} 1 & \cellcolor[RGB]{208,240,192} 130 & \cellcolor[RGB]{208,240,192} 0.615 & \cellcolor[RGB]{208,240,192} 46.00 & \cellcolor[RGB]{208,240,192} 16 & \cellcolor[RGB]{208,240,192} 1 & \cellcolor[RGB]{208,240,192} 130 & 0.593 & \cellcolor[RGB]{208,240,192} 46.00 & \cellcolor[RGB]{208,240,192} 16 & \cellcolor[RGB]{208,240,192} 1 & \cellcolor[RGB]{208,240,192} 130 & 0.593 & \cellcolor[RGB]{208,240,192} 46.00	\NN
	RAM & \cellcolor[RGB]{208,240,192} 79 & \cellcolor[RGB]{208,240,192} 40 & \cellcolor[RGB]{208,240,192} 294 & \cellcolor[RGB]{208,240,192} 0.225 & 22.0 & 135 & 57 & 443 & 0.201 & \cellcolor[RGB]{208,240,192} 21.00 & 1464 & 815 & 5824 & -0.159 & 164.0	\NN
	accu & 15 & 26 & 79 & \cellcolor[RGB]{208,240,192} 0.174 & \cellcolor[RGB]{208,240,192} 18.00 & \cellcolor[RGB]{208,240,192} 12 & \cellcolor[RGB]{208,240,192} 14 & \cellcolor[RGB]{208,240,192} 59 & 0.163 & 19.0 & \cellcolor[RGB]{208,240,192} 12 & \cellcolor[RGB]{208,240,192} 14 & 60 & 0.073 & 19.0	\NN
	adder\_16bit & 16 & 30 & 103 & N/A & 1.46e-07 & 15 & 28 & 98 & N/A & 1.38e-07 & \cellcolor[RGB]{208,240,192} 9 & \cellcolor[RGB]{208,240,192} 18 & \cellcolor[RGB]{208,240,192} 50 & N/A & \cellcolor[RGB]{208,240,192} 4.9e-08	\NN
	adder\_32bit & 60 & 105 & 306 & N/A & 4.58e-07 & - & - & - & - & - & \cellcolor[RGB]{208,240,192} 35 & \cellcolor[RGB]{208,240,192} 63 & \cellcolor[RGB]{208,240,192} 194 & N/A & \cellcolor[RGB]{208,240,192} 2.8e-07	\NN
	adder\_8bit & 21 & 38 & 87 & N/A & 9.7e-08 & \cellcolor[RGB]{208,240,192} 8 & \cellcolor[RGB]{208,240,192} 12 & \cellcolor[RGB]{208,240,192} 45 & N/A & \cellcolor[RGB]{208,240,192} 6.8e-08 & 21 & 38 & 87 & N/A & 9.7e-08	\NN
	adder\_pipe\_64bit & \cellcolor[RGB]{208,240,192} 117 & \cellcolor[RGB]{208,240,192} 71 & \cellcolor[RGB]{208,240,192} 514 & \cellcolor[RGB]{208,240,192} 0.062 & \cellcolor[RGB]{208,240,192} 127.00 & - & - & - & - & - & - & - & - & - & -	\NN
	alu & \cellcolor[RGB]{208,240,192} 404 & \cellcolor[RGB]{208,240,192} 646 & \cellcolor[RGB]{208,240,192} 2153 & N/A & \cellcolor[RGB]{208,240,192} 2.7e-06 & - & - & - & - & - & - & - & - & - & -	\NN
	asyn\_fifo & \cellcolor[RGB]{208,240,192} 27 & \cellcolor[RGB]{208,240,192} 30 & \cellcolor[RGB]{208,240,192} 120 & \cellcolor[RGB]{208,240,192} -1.331 & \cellcolor[RGB]{208,240,192} 29.00 & - & - & - & - & - & - & - & - & - & -	\NN
	calendar & \cellcolor[RGB]{208,240,192} 16 & \cellcolor[RGB]{208,240,192} 22 & \cellcolor[RGB]{208,240,192} 90 & \cellcolor[RGB]{208,240,192} 0.256 & 22.0 & - & - & - & - & - & 18 & 23 & 93 & 0.147 & \cellcolor[RGB]{208,240,192} 21.00	\NN
	counter\_12 & \cellcolor[RGB]{208,240,192} 2 & \cellcolor[RGB]{208,240,192} 4 & \cellcolor[RGB]{208,240,192} 15 & \cellcolor[RGB]{208,240,192} 0.537 & \cellcolor[RGB]{208,240,192} 13.00 & \cellcolor[RGB]{208,240,192} 2 & \cellcolor[RGB]{208,240,192} 4 & \cellcolor[RGB]{208,240,192} 15 & \cellcolor[RGB]{208,240,192} 0.537 & \cellcolor[RGB]{208,240,192} 13.00 & \cellcolor[RGB]{208,240,192} 2 & \cellcolor[RGB]{208,240,192} 4 & \cellcolor[RGB]{208,240,192} 15 & \cellcolor[RGB]{208,240,192} 0.537 & \cellcolor[RGB]{208,240,192} 13.00	\NN
	div\_16bit & \cellcolor[RGB]{208,240,192} 315 & \cellcolor[RGB]{208,240,192} 575 & \cellcolor[RGB]{208,240,192} 1516 & N/A & \cellcolor[RGB]{208,240,192} 1.8e-06 & - & - & - & - & - & - & - & - & - & -	\NN
	edge\_detect & \cellcolor[RGB]{208,240,192} 2 & \cellcolor[RGB]{208,240,192} 2 & \cellcolor[RGB]{208,240,192} 7 & \cellcolor[RGB]{208,240,192} 0.634 & \cellcolor[RGB]{208,240,192} 13.00 & \cellcolor[RGB]{208,240,192} 2 & 3 & 10 & 0.444 & \cellcolor[RGB]{208,240,192} 13.00 & \cellcolor[RGB]{208,240,192} 2 & \cellcolor[RGB]{208,240,192} 2 & \cellcolor[RGB]{208,240,192} 7 & \cellcolor[RGB]{208,240,192} 0.634 & \cellcolor[RGB]{208,240,192} 13.00	\NN
	freq\_div & 10 & 18 & 42 & \cellcolor[RGB]{208,240,192} 0.255 & \cellcolor[RGB]{208,240,192} 13.00 & \cellcolor[RGB]{208,240,192} 9 & \cellcolor[RGB]{208,240,192} 17 & \cellcolor[RGB]{208,240,192} 37 & \cellcolor[RGB]{208,240,192} 0.255 & \cellcolor[RGB]{208,240,192} 13.00 & 10 & 18 & 42 & \cellcolor[RGB]{208,240,192} 0.255 & \cellcolor[RGB]{208,240,192} 13.00	\NN
	fsm & \cellcolor[RGB]{208,240,192} 5 & \cellcolor[RGB]{208,240,192} 8 & \cellcolor[RGB]{208,240,192} 23 & \cellcolor[RGB]{208,240,192} 0.371 & \cellcolor[RGB]{208,240,192} 12.00 & - & - & - & - & - & \cellcolor[RGB]{208,240,192} 5 & \cellcolor[RGB]{208,240,192} 8 & \cellcolor[RGB]{208,240,192} 23 & 0.348 & \cellcolor[RGB]{208,240,192} 12.00	\NN
	multi\_16bit & 100 & \cellcolor[RGB]{208,240,192} 123 & 547 & -0.526 & \cellcolor[RGB]{208,240,192} 41.00 & - & - & - & - & - & \cellcolor[RGB]{208,240,192} 94 & 146 & \cellcolor[RGB]{208,240,192} 505 & \cellcolor[RGB]{208,240,192} -0.360 & 45.0	\NN
	multi\_booth\_8bit & \cellcolor[RGB]{208,240,192} 48 & \cellcolor[RGB]{208,240,192} 89 & \cellcolor[RGB]{208,240,192} 300 & \cellcolor[RGB]{208,240,192} -1.702 & \cellcolor[RGB]{208,240,192} 54.00 & - & - & - & - & - & 60 & 102 & 319 & -1.862 & 55.0	\NN
	multi\_pipe\_4bit & \cellcolor[RGB]{208,240,192} 10 & \cellcolor[RGB]{208,240,192} 18 & \cellcolor[RGB]{208,240,192} 50 & \cellcolor[RGB]{208,240,192} 0.343 & \cellcolor[RGB]{208,240,192} 16.00 & - & - & - & - & - & 17 & 31 & 58 & -0.036 & 17.0	\NN
	multi\_pipe\_8bit & 58 & \cellcolor[RGB]{208,240,192} 89 & 215 & -0.061 & \cellcolor[RGB]{208,240,192} 31.00 & - & - & - & - & - & \cellcolor[RGB]{208,240,192} 54 & \cellcolor[RGB]{208,240,192} 89 & \cellcolor[RGB]{208,240,192} 213 & \cellcolor[RGB]{208,240,192} -0.026 & \cellcolor[RGB]{208,240,192} 31.00	\NN
	parallel2serial & \cellcolor[RGB]{208,240,192} 4 & \cellcolor[RGB]{208,240,192} 7 & \cellcolor[RGB]{208,240,192} 17 & \cellcolor[RGB]{208,240,192} 0.612 & \cellcolor[RGB]{208,240,192} 13.00 & - & - & - & - & - & - & - & - & - & -	\NN
	pe & \cellcolor[RGB]{208,240,192} 23 & \cellcolor[RGB]{208,240,192} 46 & \cellcolor[RGB]{208,240,192} 232 & \cellcolor[RGB]{208,240,192} -0.265 & \cellcolor[RGB]{208,240,192} 52.00 & \cellcolor[RGB]{208,240,192} 23 & \cellcolor[RGB]{208,240,192} 46 & \cellcolor[RGB]{208,240,192} 232 & \cellcolor[RGB]{208,240,192} -0.265 & \cellcolor[RGB]{208,240,192} 52.00 & \cellcolor[RGB]{208,240,192} 23 & \cellcolor[RGB]{208,240,192} 46 & \cellcolor[RGB]{208,240,192} 232 & \cellcolor[RGB]{208,240,192} -0.265 & \cellcolor[RGB]{208,240,192} 52.00	\NN
	pulse\_detect & \cellcolor[RGB]{208,240,192} 3 & \cellcolor[RGB]{208,240,192} 5 & \cellcolor[RGB]{208,240,192} 17 & \cellcolor[RGB]{208,240,192} 0.559 & \cellcolor[RGB]{208,240,192} 12.00 & - & - & - & - & - & - & - & - & - & -	\NN
	radix2\_div & \cellcolor[RGB]{208,240,192} 61 & \cellcolor[RGB]{208,240,192} 97 & \cellcolor[RGB]{208,240,192} 313 & \cellcolor[RGB]{208,240,192} -0.123 & \cellcolor[RGB]{208,240,192} 28.00 & - & - & - & - & - & - & - & - & - & -	\NN
	right\_shifter & \cellcolor[RGB]{208,240,192} 3 & \cellcolor[RGB]{208,240,192} 0 & \cellcolor[RGB]{208,240,192} 16 & \cellcolor[RGB]{208,240,192} 0.637 & \cellcolor[RGB]{208,240,192} 16.00 & \cellcolor[RGB]{208,240,192} 3 & \cellcolor[RGB]{208,240,192} 0 & \cellcolor[RGB]{208,240,192} 16 & \cellcolor[RGB]{208,240,192} 0.637 & \cellcolor[RGB]{208,240,192} 16.00 & \cellcolor[RGB]{208,240,192} 3 & \cellcolor[RGB]{208,240,192} 0 & \cellcolor[RGB]{208,240,192} 16 & \cellcolor[RGB]{208,240,192} 0.637 & \cellcolor[RGB]{208,240,192} 16.00	\NN
	serial2parallel & 10 & 7 & 41 & 0.309 & 17.0 & \cellcolor[RGB]{208,240,192} 6 & \cellcolor[RGB]{208,240,192} 4 & \cellcolor[RGB]{208,240,192} 27 & \cellcolor[RGB]{208,240,192} 0.591 & \cellcolor[RGB]{208,240,192} 16.00 & 8 & 6 & 35 & 0.391 & \cellcolor[RGB]{208,240,192} 16.00	\NN
	signal\_generator & \cellcolor[RGB]{208,240,192} 6 & \cellcolor[RGB]{208,240,192} 6 & 30 & \cellcolor[RGB]{208,240,192} 0.559 & \cellcolor[RGB]{208,240,192} 14.00 & \cellcolor[RGB]{208,240,192} 6 & \cellcolor[RGB]{208,240,192} 6 & \cellcolor[RGB]{208,240,192} 27 & 0.543 & \cellcolor[RGB]{208,240,192} 14.00 & \cellcolor[RGB]{208,240,192} 6 & \cellcolor[RGB]{208,240,192} 6 & 30 & \cellcolor[RGB]{208,240,192} 0.559 & \cellcolor[RGB]{208,240,192} 14.00	\NN
	synchronizer & - & - & - & - & - & \cellcolor[RGB]{208,240,192} 20 & \cellcolor[RGB]{208,240,192} 0 & \cellcolor[RGB]{208,240,192} 18 & \cellcolor[RGB]{208,240,192} -1.207 & \cellcolor[RGB]{208,240,192} 22.00 & \cellcolor[RGB]{208,240,192} 20 & \cellcolor[RGB]{208,240,192} 0 & \cellcolor[RGB]{208,240,192} 18 & \cellcolor[RGB]{208,240,192} -1.207 & \cellcolor[RGB]{208,240,192} 22.00	\NN
	traffic\_light & 18 & 31 & 99 & 0.149 & \cellcolor[RGB]{208,240,192} 18.00 & - & - & - & - & - & \cellcolor[RGB]{208,240,192} 17 & \cellcolor[RGB]{208,240,192} 25 & \cellcolor[RGB]{208,240,192} 87 & \cellcolor[RGB]{208,240,192} 0.218 & \cellcolor[RGB]{208,240,192} 18.00	\NN
	width\_8to16 & 14 & \cellcolor[RGB]{208,240,192} 3 & 52 & 0.372 & \cellcolor[RGB]{208,240,192} 21.00 & \cellcolor[RGB]{208,240,192} 13 & \cellcolor[RGB]{208,240,192} 3 & \cellcolor[RGB]{208,240,192} 51 & \cellcolor[RGB]{208,240,192} 0.377 & \cellcolor[RGB]{208,240,192} 21.00 & 14 & \cellcolor[RGB]{208,240,192} 3 & 53 & 0.352 & \cellcolor[RGB]{208,240,192} 21.00	\ML
	\# Best Quality & 17/28 & 21/28 & 17/28 & 18/23 & 22/28 & 12/14 & 11/14 & 11/14 & 7/12 & 12/14 & 14/22 & 14/22 & 12/22 & 10/19 & 16/22	\LL
\end{tabular}
}
\end{table*}

\subsection{Discussion}
\subsubsection{\textbf{Ablation Study}}
To evaluate the effectiveness of our designed self-correction and self-verification mechanisms, we further conduct an ablation study to evaluate the performance of GPT-4 when equipped solely with self-verification~(generating a test bench and conducting code walk-throughs) and its performance with only self-correction~(revising code based on simulation feedback). We show the results in Table~\ref{tab:passrate}. From the results, we observe the effectiveness of both self-verification and self-correction, each contributing to improvements over the GPT-4 baseline. GPT-4 with self-verification achieves comparable and even slightly better results than GPT-4 with self-correction. This suggests that the process of the code walking through is crucial for the model’s deeper understanding of the code and the task, thereby aiding in code enhancement. Furthermore, having a valid test bench to obtain real feedback from simulations that directly pinpoints specific areas in need of refinement, also contributes to code refinement.

\begin{table}[t]
    \caption{The quality of the generated test bench during the self-verification step and the functionality correctness of the corresponding RTL code on the RTLLM benchmark.}
    \label{tab:testbench_quality}
    \centering
    \resizebox{\linewidth}{!}{
    \begin{tabular}{|c|c|c|} \hline 
    Test bench quality     &   \#test bench&\#correct RTL (correctness rate)\\ \hline 
 fully correct& 11 & 10 (90.0\%)\\ \hline 
         correct test cases&  12    &9 (75.0\%)\\ \hline 
 incorrect& 6&3 (50.0\%)\\ \hline\end{tabular}}

\end{table}

\begin{table}[h]
\centering
\caption{Evaluation on the robustness of \nickname{}'s RTL code generation with GPT-4.}
\label{tab:llm_consistency}
\begin{tabular}{|c|c|c|}
\hline
Benchmark & Avg pass@5 & std \\ \hline
  VerilogEval-Human                 &         0.626         &  0.007                  \\ \hline
  VerilogEval-Machine                    &          0.780          &   0.008                    \\ \hline
RTLLM                   & 0.752                   &     0.289                   \\ \hline
\end{tabular}

\end{table}

\subsubsection{\textbf{Quality of the Generated Test Benches}}\label{sec:effect_self_veri}
Although test bench generation is a side task to help improve the RTL code generation, it is itself a kind of code generation task. 
Thus we study the quality of the test bench generation by manually evaluating the 29 test benches generated during the self-verification process on the RTLLM benchmark. We categorize them into three quality levels: (1) fully correct, where the test bench is both syntactically and semantically accurate, providing correct test cases and timing settings;
(2) correct test cases, where the test bench contains some bugs but includes accurate input and output cases, which are crucial for effective code walk-throughs; and (3) incorrect, where the test bench fails to include accurate input and output pairs but still provides some input cases and timing setting for walk-throughs.

Table~\ref{tab:testbench_quality} illustrates the relationship between the quality of generated test benches and the correctness of the corresponding RTL code. Among 11 generated test benches deemed fully correct, 10 resulted in correct RTL code, illustrating that a high-quality test bench enhances the self-verification process and leads to a high rate of RTL correctness (90.9\%). 
Test benches categorized as having correct test cases achieve a 75\% correctness rate. In contrast, test benches that are incorrect lead to only 50\% correct RTL code. Notably, \nickname{} is able to generate 23 out of 29 test benches that include at least correct test cases, which contribute to the majority of the correct RTL code. 

Our study underscores that test bench generation is essential for RTL design but poses significant challenges as a code generation task. It demands not only correct input and output cases but also precise timing requirements to adequately cover corner cases or critical paths. Improving test bench generation to suggest high-quality test bench remains a key area for our future research.



\subsubsection{\textbf{Robustness of \nickname{}}}
To evaluate the robustness of \nickname{}, we further evaluate whether the proposed self-correction and self-verification mechanisms work well consistently across various prompt variants. For the initial prompt, self-verification prompt, and self-correction prompt, we craft 5 prompt variants for each of them that slightly vary in their wording but maintain the same fundamental task. For example, in the initial prompt, variants could range from "outline the steps needed before generating RTL code" to "plan the solutions step-by-step then implement accordingly". Then in each testing round, we randomly select one variant from each set of prompts to ensure that each test experiences a slightly different input context, while still following the standard \nickname{} workflow. This randomness allows us to examine how minor changes in prompts affect the consistency of the LLM's output and the robustness of \nickname{}. 
We repeat the test five times and then calculate the average score and the standard deviation (std), as reported in Table~\ref{tab:llm_consistency}. The results indicate that the scores are very stable with a minimal standard deviation, demonstrating that the effectiveness of \nickname{} workflow is consistent across various wordings in prompts. During testing, we observed that different combinations of prompt variants lead to various code generation and refinement traces. However, with the proposed \nickname{} framework, the model is able to generate diverse, yet correct, code with consistently high pass rate scores.

\section{Conclusion}
We propose \textit{\nickname{}}, an LLM-powered assistant for Verilog RTL design, which suggests high-quality RTL code with test benches. \nickname{} is an automatic prompting system, integrating with the simulator to enable LLMs to iteratively perform self-correction and self-verification to improve the quality of generated RTL code. Comprehensive evaluation across various benchmarks shows significant improvements in both syntax and functionality correctness over existing LLM implementations for RTL code generation, establishing \nickname{}'s potential efficacy in reducing the need for human intervention and making RTL design more accessible to novices. Future work will focus on integrating synthesis results into the feedback loop of \nickname{} to improve the performance of the generated RTL code. Additionally, generating reliable test benches to effectively cover critical cases and assist test bench design is an essential aspect to explore in assisting RTL design.

\bibliographystyle{ACM-Reference-Format}
\bibliography{sample-base}

\end{document}